\documentstyle[12pt,times,epsf,psfig]{article}

\begin{document}

\baselineskip 6mm
\renewcommand{\thefootnote}{\fnsymbol{footnote}}


\newcommand{\nc}{\newcommand}
\newcommand{\rnc}{\renewcommand}


\rnc{\baselinestretch}{1.24}	
\setlength{\jot}{6pt} 		
\rnc{\arraystretch}{1.24}   	

\makeatletter
\@addtoreset{equation}{section}
\makeatother                      



\nc{\be}{\begin{equation}}
\nc{\ee}{\end{equation}}
\nc{\bea}{\begin{eqnarray}}
\nc{\eea}{\end{eqnarray}}
\nc{\xx}{\nonumber\\}

\nc{\eq}[1]{(\ref{#1})}



\nc{\np}[3]{Nucl. Phys. {\bf B#1} (#2) #3}
\nc{\pl}[3]{Phys. Lett. {\bf #1B} (#2) #3}
\nc{\prl}[3]{Phys. Rev. Lett.{\bf #1} (#2) #3}
\nc{\prd}[3]{Phys. Rev. {\bf D#1} (#2) #3}
\nc{\ap}[3]{Ann. Phys. {\bf #1} (#2) #3}
\nc{\prep}[3]{Phys. Rep. {\bf #1} (#2) #3}
\nc{\rmp}[3]{Rev. Mod. Phys. {\bf #1} (#2) #3}
\nc{\cmp}[3]{Comm. Math. Phys. {\bf #1} (#2) #3}
\nc{\mpl}[3]{Mod. Phys. Lett. {\bf #1} (#2) #3}
\nc{\cqg}[3]{Class. Quant. Grav. {\bf #1} (#2) #3}
\nc{\jhep}[3]{J. High Energy Phys. {\bf #1} (#2) #3}


\def\vare{\varepsilon}
\def\bz{\bar{z}}
\def\bw{\bar{w}}


\def\CA{{\cal A}}
\def\CC{{\cal C}}
\def\CD{{\cal D}}
\def\CE{{\cal E}}
\def\CF{{\cal F}}
\def\CG{{\cal G}}
\def\CI{{\cal I}}
\def\CT{{\cal T}}
\def\CM{{\cal M}}
\def\CN{{\cal N}}
\def\CP{{\cal P}}
\def\CL{{\cal L}}
\def\CV{{\cal V}}
\def\CS{{\cal S}}
\def\CW{{\cal W}}
\def\CY{{\cal Y}}
\def\CS{{\cal S}}
\def\CO{{\cal O}}
\def\CP{{\cal P}}
\def\CN{{\cal N}}
\def\CZ{{\cal Z}}


\def\IR{{\hbox{{\rm I}\kern-.2em\hbox{\rm R}}}}
\def\IB{{\hbox{{\rm I}\kern-.2em\hbox{\rm B}}}}
\def\IN{{\hbox{{\rm I}\kern-.2em\hbox{\rm N}}}}
\def\IC{\,\,{\hbox{{\rm I}\kern-.59em\hbox{\bf C}}}}
\def\IZ{{\hbox{{\rm Z}\kern-.4em\hbox{\rm Z}}}}
\def\IP{{\hbox{{\rm I}\kern-.2em\hbox{\rm P}}}}
\def\IH{{\hbox{{\rm I}\kern-.4em\hbox{\rm H}}}}
\def\ID{{\hbox{{\rm I}\kern-.2em\hbox{\rm D}}}}


\def\a{\alpha}
\def\b{\beta}
\def\ga{\gamma}
\def\d{\delta}
\def\ep{\epsilon}
\def\ph{\phi}
\def\k{\kappa}
\def\l{\lambda}
\def\m{\mu}
\def\n{\nu}
\def\th{\theta}
\def\rh{\rho}
\def\s{\sigma}
\def\t{\tau}
\def\w{\omega}
\def\G{\Gamma}


\def\half{\frac{1}{2}}
\def\imp{\Longrightarrow}
\def\dint#1#2{\int\limits_{#1}^{#2}}
\def\goto{\rightarrow}
\def\para{\parallel}
\def\brac#1{\langle #1 \rangle}
\def\del{\nabla}
\def\grad{\nabla}
\def\curl{\nabla\times}
\def\div{\nabla\cdot}
\def\p{\partial}
\def\e{\epsilon_0}


\def\Tr{{\rm Tr}}
\def\det{{\rm det}}
\def\im{{\rm Im~}}
\def\re{{\rm Re~}}


\def\Kahler{K\"{a}hler}


\def\e{\varepsilon}
\def\bA{\bar{A}}
\def\c{\zeta}

\begin{titlepage}
\hfill\parbox{4cm}
{hep-th/0011119}

\vspace{15mm}
\centerline{\Large \bf Open string derivation of 
winding states} \vspace{5mm} \centerline{\Large \bf  
in thermal noncommutative field theories}
\vspace{10mm}
\begin{center}  
Youngjai Kiem\footnote{ykiem@newton.skku.ac.kr},
Dong Hyun Park\footnote{donghyun@newton.skku.ac.kr},
and Haru-Tada Sato\footnote{haru@taegeug.skku.ac.kr}\\[2mm] 
{\sl BK21 Physics Research Division and Institute of Basic Science, 
Sungkyunkwan University, 
Suwon 440-746, Korea}
\end{center}

\thispagestyle{empty}
\vskip 40mm

\centerline{\bf ABSTRACT}
\vskip 5mm
\noindent
The `winding state' behavior appears in the two-loop nonplanar 
contribution to the partition function in thermal noncommutative 
field theories.  We derive this feature directly from the purely 
open string theory analysis in the presence of the constant 
background $B$-field; we compute the two-loop partition function 
for worldsheets with a handle and a boundary when the time 
direction of the Euclideanized target space is compactified.  
In contrast to the closed-string-inspired approach, it is 
not necessary to add infinite number of extra degrees of freedom.  
Furthermore, we find a piece of supporting evidence toward the 
conjecture that, in the UV limit, the noncommutativity parameter 
plays the role of the effective string scale in noncommutative 
field theories.  
\vspace{2cm}
\end{titlepage}

\baselineskip 7mm
\renewcommand{\thefootnote}{\arabic{footnote}}
\setcounter{footnote}{0}

Open string theories in the presence of the constant background
(spatial) NS-NS two-form gauge field ($B$) \cite{string}, 
describing the dynamics of noncommutative D$p$-branes, reduce 
to $(p+1)$-dimensional noncommutative field theories 
upon taking a decoupling limit, where the string length scale
$\alpha^{\prime} \rightarrow 0$ \cite{witten}.  
The resulting noncommutative field theories are known to possess 
remarkably stringy features such as UV/IR mixing \cite{uvir} and 
Morita equivalence (related to string dualities) \cite{chull}.
Generally, when there are compact directions in the target space, 
the behavior of commutative point particle field theories 
are markedly different from that of string theories.  
Considering the similar situation in 
noncommutative field theories, however, might reveal more
of stringy features.  One such example is the `winding state'
behavior observed in thermal noncommutative field theories \cite{fisch}.
Formally, thermal field theories are obtained by Euclideanizing
the time direction and compactifying it with the period 
given by the inverse temperature.  To understand this feature
from the string theory point of view, one naturally has to
consider a compact direction in the target space, which is
parallel to the D$p$-branes.

In general, there are three kinds of target 
space compactifications that can be of interests.  The first
kind is the compactification of the directions perpendicular
to noncommutative D-branes.  Especially in the context of
noncommutative warped compactifications \cite{verlinde}, 
this can be of physical importance.  
The second kind is the compactification
of the directions parallel to the D-branes and to the 
nonzero $B$ fields, which has also been
actively pursued \cite{mehen}.  Our main interest in this
note is the third kind, where one compactifies a target
space direction parallel to the D-branes and perpendicular 
to the nonzero $B$ field directions; we analyze the nonplanar 
two-loop partition function of string theory with the 
constant and spatial $B$ field when such kind of 
compactification is present. 
Being the nonplanar vacuum diagram, the 
worldsheets in consideration have a handle ($g=1$) and a 
boundary ($b=1$).  We show that the `winding state' behavior
observed in thermal noncommutative field theories \cite{fisch} 
can be reproduced from the string theory computations
on $g=1$, $b=1$ worldsheets upon taking the decoupling
limit $\alpha^{\prime} \rightarrow 0$.  An interesting
point is that even the decoupled theories, i.e., noncommutative 
field theories, turn out to be very stringy; our 
perturbative string theory 
calculations are consistent with the proposal \cite{kek} that, in 
noncommutative field theories, the noncommutativity parameter 
$\theta$, that appears in the $*$-product between operators 
\begin{equation}
 \phi_1 (x) * \phi_2 (x) = \exp \left( \frac{i}{2} 
\theta^{\mu \nu} \frac{\partial}{\partial y^{\mu}}
 \frac{\partial}{\partial z^{\nu}} \right)
  \phi_1 (y ) \phi_2 (z) |_{y=z=x} ~~~ ,
\end{equation}
plays the role of the 
effective string scale $\alpha^{\prime}_{eff}$.    

The way that we derive the `winding states' in the $\phi^4$
noncommutative scalar field theory is based on {\em purely}
open string analysis combined with the stretched string
interpretation of \cite{liu}.  It is instructive to compare
our approach to that of \cite{arcioni} based on closed string
type approach.  Due to the dipole effect advocated 
in \cite{suss}, which lies at the heart of the stretched string 
interpretation of \cite{liu}, an open string in the presence
of the constant $B$ field tends to stretch as it gains 
momentum, indicating the nonlocal nature of noncommutative
field theories.  The UV/IR mixing \cite{uvir}, the appearance of 
the IR divergence from the UV regime of the loop momentum integration
in nonplanar amplitudes, can be explained by such stretched 
strings.  Typically, the stretching length of an open string 
acts as the UV regulator of the nonplanar amplitudes.  Since
the stretching length is proportional to the momentum, the
amplitudes tend to diverge as the UV length cutoff goes to
zero size in the zero momentum limit.  As we will see shortly, 
this UV/IR mixing effect plays a crucial role in the physics of
`winding states'.  As inspired by Seiberg, Raamsdonk and
Minwalla \cite{uvir}, however, one might 
attempt to add extra `closed string'-like 
degrees of freedom, which, upon integrating them out, might explain 
such nonlocal behaviors as the UV/IR mixing and the `winding
states'.  Such attempts reported in the literature have one
feature in common; we have to add infinite number of extra
degrees of freedom.  In the context of `winding states', one
has to consider an infinite number of modes that resemble 
the closed string winding modes \cite{arcioni}.  
In the context of the UV/IR mixing phenomenon, one can argue 
that the full massive closed string corrections cannot be 
neglected in the closed string channel description \cite{rajaraman}.  
In contrast, in the {\rm purely}
open string approaches for the same problems, as we explicitly
verify for the `winding state' issue here, one does not 
have to add infinite extra degrees of freedom; in fact, one
finds that the conventional quantized momenta (loop momenta
in the particular case in consideration) of the open
string zero mode parts in the compact direction conspire to 
produce the effective `winding states'.  This is in line with
the conventional wisdom that the dynamics that is easily described 
by the low energy degrees of freedom in one theory (open strings) has 
rather complicated dual theory descriptions (in terms of
closed strings)\cite{shenker}.  

Our analysis starts from the computation of the partition
functions on $(gb)$ worldsheets with $g$ handles and $b$
boundaries in the presence of the constant background
$B$ field and a compact direction perpendicular to it.  
When there is the constant background $B$ field, the open 
string metric $G^{\mu \nu}$ and the noncommutativity
parameter $\theta^{\mu \nu}$ are related to the corresponding
closed string quantities via
\begin{equation}
 G^{\mu \nu} = (g_{\mu \nu} + B_{\mu \nu})^{-1}_S  ~~~ , ~~~
 \theta^{\mu \nu} =  2 \pi \alpha^{\prime} ( g_{\mu \nu} +
  B_{\mu \nu} )^{-1}_A ~,
\end{equation}
where the subscripts $S$ and $A$ denote the symmetric and
the antisymmetric parts of a matrix, respectively. 
In the absence of the compact direction, the computation
of the partition functions was performed in \cite{technique,recent}
and we will use their results in this note.  The new element
here will be the incorporation of the topological sectors
resulting from the existence of the compact direction.

For the description of worldsheets, it is convenient to first 
consider the $(g0)$ worldsheets.  On a $(g0)$ worldsheet, 
there are $2g$ homology cycles forming a basis, $a_{\alpha}$ and 
$b_{\alpha}$ ($\alpha = 1 , \cdots , g$) with canonical
intersection parings, and $2g$ Abelian differentials 
$\omega_{\alpha}$ (holomorphic) and $\bar{\omega}_{\alpha}$ 
(antiholomorphic).   These Abelian differentials are
normalized along the $a_{\alpha}$-cycles and, when 
integrated over $b_{\alpha}$ cycles, determine the
$g \times g$ period matrix $\tau$ 
\begin{equation}
 \oint_{a_{\alpha}} \omega_{\beta} = \delta_{\alpha \beta} ~~~ , ~~~
 \oint_{b_{\alpha}} \omega_{\beta} = \tau_{\alpha \beta} ~ .
\end{equation}
Up to three loops, it is known that the moduli space of
the worldsheets are parameterized by the symmetric 
period matrix without any redundancy.  In this closed
string setup, it is well known how to incorporate the effects
caused by a compact target space direction \cite{atick}.
We find it useful to review the conventional derivations
of such effects.  The Euclideanized string partition function in 
the path integral formalism contains a factor 
\begin{equation}
\int \!{\cal{D}} \psi {\cal{D}}\tau \,{\cal{D}}h \,{\cal{D}}X 
\exp \left(  - S_X \right) \times \cdots \label{partition} ~ ,
\end{equation}
where ${\cal{D}} \psi$, ${\cal{D}}\tau$, ${\cal{D}}h$ and 
${\cal{D}}X$ represent the ghost, the moduli, 
the worldsheet metric and the target space coordinate $X$ 
integrations.
The action $S_X$ in conformal gauge is given by
\begin{equation}
S=\frac{1}{4 \pi \alpha^{\prime}} 
\int\! d^2 \sigma \sqrt{h}\, h^{\alpha\beta} \partial_\alpha
X \partial_\beta X  = \frac{i}{2 \pi \alpha^{\prime}}
 \int \! dz\! \wedge\! d\bar{z}\, \partial X
\bar{\partial}  X ~ . \label{action}
\end{equation}
Here we have introduced the local holomorphic and antiholomorphic
coordinates $z$ and $\bar{z}$.
We assume that $X$ is compactified to a circle with a radius $R$ via 
the identification $X \simeq X + 2 \pi R$.  The topological sector
of $X$ field can be written as
\begin{equation}
dX = p^\alpha \omega_\alpha +\bar{p}^\alpha \bar{\omega}_\alpha +
 {\rm fluctuation ~ modes} ~ , \label{quan} 
\end{equation}
where $p^\alpha$ and $\bar{p}^\alpha$ are independent holomorphic
and antiholomorphic zero mode parameters for closed strings (modulo
the overall level matching conditions).  These parameters are 
determined by the holonomy properties under the shift along
each cycle:
\begin{eqnarray}
\oint_{a_\alpha}\!\! d X &=&  p_\alpha
+\bar{p}_\alpha =2 \pi R n_\alpha \nonumber \\
\oint_{b_\alpha}\!\! d X &=&  
\tau_{\alpha\beta} p^\beta + \bar{\tau}_{\alpha\beta} \bar{p}^\beta
=2 \pi R m_\alpha \label{wind} ~ ,
\end{eqnarray}
where $n_{\alpha}$ and $m_{\alpha}$ are integers.  Once these
are determined, the topological sector contribution to the 
partition function can be computed by evaluating the action 
Eq.~(\ref{action})
\begin{equation}
 \CZ_t^{(g) c} = \sqrt{\kappa }
   \sum_{n_{\alpha}, ~ m_{\alpha}} \exp \Big[ 
- \kappa \Big( m_\alpha\ ( \im  \tau )^{-1 ~ \alpha\beta} 
m_\beta - 2 m_\alpha ( ( \im \tau )^{-1} {\rm Re}\;\tau)^{\alpha\beta}
n_\beta \label{ft1}
\end{equation}
\[ + n_{\alpha} (  \bar{\tau} ( \im \tau )^{-1} \tau )^{\alpha \beta}
  n_{\beta} \Big) \Big] ~ , \]
by using the formula that holds for a worldsheet without boundaries
\begin{equation}
\int \! dz\! \wedge\! d\bar{z}\, \partial X
\bar{\partial} X = \sum_\alpha \left( \oint_{a_\alpha}\!\! dz \partial X
\oint_{b_\beta}\!\! d\bar{z} \bar{\partial} X - \oint_{a_\alpha}\!\! d\bar{z}
\bar{\partial} X \oint_{b_\beta}\!\! dz \partial X \right) ~  ,
\label{wow}
\end{equation}
and $\kappa = \pi R^2 / \alpha^{\prime}$.  The multiplicative
factor $\sqrt{\kappa}$ in Eq.~(\ref{ft1}) comes from the 
center-of-mass part functional integration.  
Hereafter and in Eq.~(\ref{ft1})
we ignore the moduli- and $\kappa$-independent multiplicative factor 
in front of the partition function.  
Using the Poisson resummation formula
\begin{equation}
\sum_{n=-\infty}^{\infty} f(n) = \sum_{m=-\infty}^{\infty}
\int_{-\infty}^{\infty} dx \, f(x) e^{2\pi i m x} ~ ,\label{prf}
\end{equation}
we can rewrite Eq.~(\ref{ft1}) as follows
\begin{equation}
  \CZ_t^{(g) c} =   \kappa^{(1-g)/2} \sqrt{ \det ~ (\im \tau )}  
 \label{ft2} 
\end{equation}
\[ \times  \sum_{m_{\alpha}, ~ n_{\alpha}}  
\exp \left[ -\kappa n_{\alpha} ~ ( {\rm Im~}\tau )^{\alpha \beta} ~ 
  n_{\beta} - \frac{\pi^2}{\kappa} m_{\alpha} ~ ( {\rm
Im~}\tau )^{\alpha \beta} ~ m_{\beta} + 2 \pi i 
 m_{\alpha}~ ( {\rm Re~}\tau )^{\alpha \beta}~ n_{\beta} 
 \right] ~ , \] 
which makes the T-duality invariance 
$R \rightarrow \alpha^{\prime} / R $ ($\kappa \rightarrow \pi^2
 / \kappa$ ),
$n_{\alpha} \rightarrow m_{\alpha}$ and $m_{\alpha} \rightarrow
 n_{\alpha}$ manifest in the exponential part.  In particular, 
for the torus with $g=1$ and $b=0$, one gets
\begin{equation}
  \CZ_t^{(1) c} = \sqrt{\tau_2} \sum_{m,n=-\infty}^{\infty}
 \exp \left[ -\kappa \tau_2 n^2 - \frac{\pi^2}{\kappa}
  \tau_2 m^2
+ 2 \pi i  \tau_1 mn \right] ~ , \label{torus}
\end{equation}
where the torus moduli $\tau = \tau_1 + i \tau_2$.

We now turn to the case of our main interest, the $g=1$ and $b=1$
worldsheets, which correspond to the nonplanar two-loop
vacuum worldsheets with the Euler characteristic $\chi = - 1$.  
A $(11)$ worldsheet can be considered as the
`folded' version of a $(20)$ worldsheet by an anticonformal 
involution $I$, the fixed points of which becoming the boundary.
As such, there are two intersecting homology cycles ${\rm a}
 = a_1 - b_2$ and ${\rm b} = b_1$ in the homology basis where the
period matrix is given by\footnote{For the conventions of
homology cycles, period matrix, etc., we follow those of
\cite{recent}.  One notes that, while the overbar of \cite{recent}
represents the action of the involution $I$, the overbar in
this note denotes the complex conjugation.} 
\begin{equation}
\tau_{\alpha\beta}=
\left( \begin{array}{cc}
iT_{11}&\frac{1}{2}+ i T_{12}\\
\frac{1}{2}+ i T_{12} & iT_{22}
\end{array}
\right)~. \label{period}
\end{equation}
This period matrix has three independent components $T_{11}$,
$T_{12}$ and $T_{22}$ corresponding to three moduli parameters
of the $(11)$ surfaces.  Of the original six moduli parameters
of the $(20)$ surfaces, the ``even'' sector satisfying the
condition $\tau = I ( \tau )$ survives the folding operation
$I$.  When compared to another
two-loop worldsheet $(03)$ with $\chi = -1$, a planar vacuum
worldsheet, that does not have intersecting homology cycles,
the key difference is the existence of the real part in 
the off-diagonal elements of the period matrix.  The analog
of Eq.~(\ref{quan}) for $(11)$ surfaces is
\begin{equation}
dX = p^\alpha \omega_\alpha + p^{\alpha} \bar{\omega}_\alpha +
 {\rm fluctuation ~ modes} ~ , \label{quan1} 
\label{zeros}
\end{equation}
where the index $\alpha$ runs over $(1,2)$.
Since the compact direction in our consideration is perpendicular
to the directions where the $B$ fields are turned on, we impose
the usual Neumann boundary condition for $X$.  This boundary
condition, in turn, sets the condition 
$p_{\alpha} = \bar{p}_{\alpha}$.  While the number of independent
$p$'s reduces from four to two, the number of independent cycles
also reduces from four to two under the folding operation
from $(20)$ surfaces.  The analog of the holonomy properties
Eqs.~(\ref{wind}) becomes
\begin{eqnarray}
\oint_{{\rm a}} ~ ~ d X  =   \oint_{a_1-b_2}\!\!\ \left( p^\alpha
\omega_\alpha + p^\alpha \bar{\omega}_\alpha \right)
=  p_1  = 2 \pi \tilde{R} n  \nonumber \\
\oint_{{\rm b}} ~ ~ d X  =    
  \oint_{b_1} \!\!\  \left( p^\alpha  
\omega_\alpha + p^\alpha \bar{\omega}_\alpha \right)
= p_2 = 2 \pi \tilde{R} m  \label{wind1} ~ ,
\end{eqnarray}
where we specified the holonomy properties under the shifts
along the ${\rm a}$ and ${\rm b}$ cycles and used the explicit
form of the period matrix, Eq.~(\ref{period}), for the evaluation
of the integrals.  Here $n$ and $m$ are integers, and $\tilde{R}$ is 
the radius of the compact direction.  The notable difference between
the nonplanar two-loop $(11)$ surfaces and the planar two-loop
$(03)$ surfaces is that, for the latter with the non-intersecting
homology cycles
$\rm{a} = b_1$ and $\rm{b} = b_2$, the holonomy properties
similar to Eqs.~(\ref{wind1}) do not constrain the values 
of $p_{\alpha}$, since the period
matrix is purely imaginary unlike the case of the 
former, Eq.~(\ref{period}). 
Inserting Eq.~(\ref{zeros}) to the open string action of the
form Eq.~(\ref{action}), we compute the topological sector
contribution to the partition function 
\begin{equation}
 \CZ_t^{(2)} = \sqrt{ \tilde{\kappa} } \sum_{n,m} \exp 
 \left[ - \tilde{\kappa} \left( T_{11}n^2 + 2T_{12}nm + 
 T_{22} m^2 \right) \right] ~ , \label{opent1}
\end{equation}
where $\tilde{\kappa} = 2 \pi \tilde{R}^2 / \alpha^{\prime}$.  
When deriving Eq.~(\ref{opent1}),
we used an identity similar to Eq.~(\ref{wow}), which is valid
for $(11)$ worldsheets,
\begin{equation}
\int \! dz \! \wedge\! d\bar{z}\, \partial X
\bar{\partial} X =  \oint_{\rm a} ~ dz ~\partial X
\oint_{\rm b} ~ d\bar{z} ~ \bar{\partial} X - \oint_{\rm a} ~ 
d\bar{z} ~ \bar{\partial} X \oint_{\rm b} ~ dz ~ \partial X  ~  ,
\label{wow2}
\end{equation}
for the $X$ satisfying the equations of motion and the
Neumann boundary condition under which the possible boundary
term contribution in Eq.~(\ref{wow2}) vanishes.  Upon using the
Poisson resummation formula Eq.~(\ref{prf}) for the $n$-summation,
we obtain 
\begin{equation}
 \CZ_t^{(2)}   = \frac{1}{\sqrt{T_{11}}}  \sum_{n,m}
\exp \left[ - \tilde{\kappa} \frac{T_{11}T_{22} -T_{12}^{~~2}}{T_{11}}m^2
-\frac{\pi^2}{\tilde{\kappa}}\frac{1}{T_{11}}n^2 - 
 2 \pi i \frac{T_{12}}{T_{11}}
 mn \right] ~ . \label{opent2}
\end{equation} 

The $(11)$ partition function along the uncompactified directions 
$X^1 , \cdots , X^p$ (including the massive mode contributions from 
the compactified direction $X^0$) when we turn on the $B$-field was 
computed in \cite{technique,recent} for noncommutative D$p$-branes.
We note that the ghost sector does not change under the influence
of the compactification and the background $B$ field \cite{technique}.
The $(11)$ partition function can be understood as resulting
from the `connected' nonplanar two-point open string insertions along
each boundary of a one-loop $(02)$ annulus.  For simplicity, we 
take $\theta_{12} = -\theta_{21} = \theta$
while setting the open string metric $G_{11} = G_{22} = 1$.  
We then have the partition function \cite{recent}
\begin{equation}
 \CZ_{\theta}^{(2)} = \sum_I \int dp_1 \int dp_2  \cdots \int dT_{11} 
\int dT_{22} \int dT_{12} ~ a_I 
\frac{|W_1 ( i T_{11} ) | }{T_{11}^{p/2}} \label{temp1}
\end{equation}
\[ \times \exp \Big[ \cdots  - 2 \pi \alpha^{\prime}
  T_{22} ( p_1 p_1 + p_2 p_2 + p_3 p_3 
\cdots + M_I^2 )  \]
\[ + 2 \pi \alpha^{\prime} \frac{T_{12}^2}{T_{11}} 
  ( p_1 p_1 + p_2 p_2 + p_3 p_3 + \cdots )
 - \frac{1}{2 \pi \alpha^{\prime}} 
 \frac{\theta^2}{4T_{11}} ( p_1 p_1 + p_2 p_2 ) 
\Big] ~ , \]    
where $W_1$ is constructed from the one-loop eta function and
the summation over $I$ goes over the intermediate
string mass states running around the connected (external)
vertex insertions.  The last term in the exponential
function of Eq.~(\ref{temp1}) is the contribution from
the stretched strings \cite{liu}, which is responsible
for the UV/IR mixing.  In Eq.~(\ref{temp1}), the two-loop moduli
parameter $T_{12}$ can be interpreted as the separation distance 
between two vertices along the imaginary
axis of the worldsheet.  The moduli $T_{22}$ corresponds
to the `length' of the connected leg between two vertex
insertions.  From \cite{technique,recent}, one notes
that Eq.~(\ref{temp1}) can be rewritten as
\begin{equation}
 \CZ_{\theta}^{(2)} = \int dT_{11} dT_{22} dT_{12} \frac{|W ( \tau ) | }
 {  \sqrt{ \det ~  ( 2 \pi \alpha^{\prime} G_{\mu \nu} \im \tau
 + \frac{i}{2} \theta_{\mu \nu} \CI ) } } ~ ,
\label{opent3}
\end{equation}
where the $\im \tau$ and intersection matrix $\CI$ are defined as 
\begin{equation}
  ( \im \tau )_{\alpha \beta} 
 = \pmatrix{ T_{11} & T_{12} \cr T_{12} & T_{22} } ~~~ , ~~~
  \CI_{\alpha \beta} = \pmatrix{ 0 & 1 \cr -1 & 0 } ~~~ , ~~~
 \det ( \im \tau ) = \ T_{11} T_{22} - T_{12}^{~~2} ~ ,  
\end{equation}
and $W( \tau ) $ is given by
\begin{equation}
  | W( \tau ) | = \prod_{a=1}^{10} 
    | \theta_a ( 0 | \tau ) |^{-2}  
\end{equation}  
and $\theta_a$'s are the ten even Riemann theta functions for the
(20) surfaces.  The indices $\alpha$ and $\beta$ run over $(1,2)$ and the
index $\mu$ and $\nu$ run over the noncompact directions $1, \cdots p$.
The full string partition function is then given by the integral
in Eq.~(\ref{opent3}) with the factor Eq.~(\ref{opent2}) inserted in the
integrand: $ \CZ^{(2)} = \CZ_{\theta}^{(2)} \CZ_t^{(2)}$.

Written in the form of Eqs.~(\ref{opent2}) and (\ref{temp1}), it is
straightforward to derive the nonplanar two-loop contribution
to the partition function of the noncommutative $\phi^4$ theory
in four dimensions.  We consider bosonic D3-branes setting 
$p=3$ in Eq.~(\ref{temp1}) and take the decoupling
limit, where we keep open string quantities such as $\theta$
and $G_{\mu \nu}$ fixed, while taking the string length scale to zero.
We also keep
\begin{equation}
 2 \pi \alpha^{\prime} T_{\alpha \beta} = t_{\alpha \beta}
\label{scale}
\end{equation}
fixed, while we take the limit $\alpha^{\prime} \rightarrow 0$.
Eq.~(\ref{scale}) is an essential scaling in the known string 
theory computations to recover the noncommutative field theory
results \cite{liu}.  In particular, this scaling exponentially 
suppresses the massive string mode 
contributions in Eq.~(\ref{temp1}) via, schematically,
$\exp ( - N_I T ) = \exp ( - ( N_I / 2 \pi \alpha^{\prime} ) t )$
for the excitation number $N_I$ state, except
the leading tachyon mass that we analytically continue to a finite 
positive $M^2$ \cite{haru}.  For the zero mode parts in 
Eq.~(\ref{temp1}), we observe that the powers of $\alpha^{\prime}$
are just right to allow
the replacement of $T$'s in the exponential function with $t$'s.
The interesting part is the topological sector in 
Eq.~(\ref{opent2}).  For the topological sector to give non-trivial 
{\em finite} contributions in the decoupling limit, we have to require 
that $\tilde{\kappa} / \alpha^{\prime}$ be kept fixed in the 
$\alpha^{\prime} \rightarrow 0$ limit, which implies that
$\tilde{R} \rightarrow 0$ in the same limit.  This consideration
leads us to define the `dual' radius $\beta$ via 
\begin{equation}
 \tilde{R} = \frac{\alpha^{\prime}}{\beta} ~~~ \rightarrow ~~~~ 
  \tilde{\kappa} = \frac{2 \pi \alpha^{\prime}}{\beta^2},
\label{scale2}
\end{equation}
where we keep $\beta$ fixed as we take the 
$\alpha^{\prime} \rightarrow 0$ limit.  The simple scaling in 
Eq.~(\ref{scale2}) has significant implications.
We note that the form of Eq.~(\ref{opent1}), which is
equivalent to Eq.~(\ref{opent2}), is precisely the summation over
the quantized loop momenta $m/ \beta$ and $n / \beta$
\begin{equation}
  \sum_{m,n} 
  \exp \left[ -   2 \pi \alpha^{\prime}  
  \pmatrix{ n / \beta \cr m / \beta}^T
 \im \tau \pmatrix{ n / \beta \cr m / \beta}  \right]
\label{haha}
\end{equation}
along the two intersecting cycles present in $(11)$ 
worldsheets \cite{technique,recent} moving in the
compact target space direction.  To summarize, since the
decoupling limit involves the $\tilde{R} \rightarrow 0$
limit, the string theory temperature should be inversed
to be the noncommutative field theory 
temperature: $T_{\rm field} = (1/T_{\rm string}) \times
 (1 / \alpha^{\prime})$.  In this
process, the `winding' mode description Eq.~(\ref{wind1}) 
naturally transmutes to the `momentum' mode 
description Eq.~(\ref{haha}).  The scaling 
Eq.~(\ref{scale2}) should be universally taken for an
arbitrary values of $p$ and low energy couplings. 
Specific to the $\phi^4$ theory, we 
further set $T_{12} = 0$ in the zero mode 
parts so that the quartic interaction vertices are produced 
in the low energy Feynman diagram description.  In
addition, we choose the
mass $M$ of the resulting particle to be very small by taking
the limit $\beta M \ll 1$, concentrating on the high temperature
regime.   Using the two integrals
\begin{equation}
 \int_0^{\infty} dt_{11} \frac{1}{t_{11}^{~~2}} \exp 
 \left( - \frac{A}{t_{11}} \right)
 = \frac{1}{A} ~~~ , ~~~ \int_0^{\infty} dt_{22} 
  \exp \left( - B t_{22} \right)
 = \frac{1}{B}  ~ ,
\end{equation}
we immediately find that
\begin{equation}
\CZ^{(2)} = - \frac{F^{(2)}}{T} = g^2 \sum_{n,m} \int dp_1 dp_2 dp_3
\frac{1}{\left( \frac{m^2}{\beta^2} + (p_1^2 + p_2^2 + p_3^2) 
  \right) \left( \pi^2 n^2 \beta^2 + \frac{1}{4} 
  \theta^2 ( p_1^2 + p_2^2 ) \right) }~, 
\label{fwinding}
\end{equation}
where $g^2$ is the coupling constant of the $\phi^4$ theory. 
The expression Eq.~(\ref{fwinding}) is precisely the noncommutative
field theory result obtained in \cite{fisch}.  

As spelled out earlier, the field theory limit partition function
shows stringy natures when $\theta \ne 0$.  
We note that the $X^1$ and $X^2$ part of the partition function
Eq.~(\ref{opent3}) contains the weight factor 
\begin{equation}
  \frac{1}{t_{11} t_{22} - t_{12}^2 + \theta^2 /4} ~.
\label{xxx}
\end{equation}
Therefore, the moduli integration over $t_{11} t_{22} - t_{12}^2$
variable has the main contribution in the regime where
$t_{11} t_{22} - t_{12}^2 \simeq \theta^2 $.  Plugging this
into Eq.~(\ref{opent2}), we find that
\begin{equation}
 \CZ^{(2)}_t \simeq \sum_{n,m}
\frac{1}{\sqrt{t_{11}}} 
\exp \left[ -  \frac{\theta^2 }{\beta^2} \frac{1}{t_{11}}m^2
- \pi^2 \beta^2 \frac{1}{t_{11}}n^2 - 
 2 \pi i \frac{t_{12}}{t_{11}}
 mn \right] ~ ,  \label{uvfact}
\end{equation} 
becoming similar to the torus partition function
Eq.~(\ref{torus}) if we identify $\tau_2 = 1/ t_{11}$ and $\tau_1 = -
t_{12}/ t_{11}$.  This fact is not surprising when one considers
the open string UV factorization channel where the boundary
of (11) surfaces shrinks to small size.  It is consistent
with the general observation that noncommutative field theories
retain the string theory topological sector 
information \cite{technique,recent}. 
Furthermore, Eq.~(\ref{uvfact}) is 
invariant under a `duality transformation'
\begin{equation}
 \beta \rightarrow \frac{1}{\pi} \frac{\theta}{ \beta}
 ~~~, ~~~ n \rightarrow m ~~~, ~~~ m \rightarrow n ~ ,
\end{equation} 
formally similar to the T-duality of closed string theory, except
that the string scale $\alpha^{\prime}$ is replaced by 
the noncommutativity parameter $\theta$.
This behavior is reminiscent of the idea advocated in \cite{kek}
that, in the noncommutative field theory, the noncommutativity
scale $\theta$ plays the role of the effective string scale
$\alpha^{\prime}$ in the UV limit.  We note that while the 
arguments of \cite{kek} are based on the dual supergravity
analysis (a strong coupling argument), our analysis
is perturbative.  The closed string T-duality is realized both
perturbatively and non-perturbatively.  Physically, 
the $\theta$ appearing in
Eq.~(\ref{xxx}) plays the role of a UV (small $t$) cutoff and 
it also corresponds to the effective stretching size of 
`loop stretched strings'.  Near the 
stretching size, or the UV cutoff length, in the moduli space, 
it appears that there exists a kind of `duality' where the 
noncommutativity
scale plays the role of the effective string scale.  In fact,
one's natural expectation is that the scale $\theta$ is the
minimum length scale that appears in the commutator
$[X^1 , X^2 ] = \theta$ \cite{string,kek}.

\section*{Acknowledgements}

We are grateful to Jin-Ho Cho, Seungjoon Hyun, Yoonbai Kim 
and Sangmin Lee for helpful discussions.  
Y.K.~would especially like to thank Chang-Yeong Lee and 
Jaemo Park for the conversations at the early stage of 
this work.  H.T.S.~was supported by the Brain Pool Program 
(KOSEF) and by the BK21 project.


\newpage

\begin{thebibliography}{99}

\bibitem{string} Y.-K. E. Cheung and M. Krogh, Nucl. Phys. {\bf B528}
           (1998) 185, hep-th/9803031; F. Ardalan, H. Arfaei, 
           M.M. Sheikh-Jabbari, JHEP 9902 (1999) 016, hep-th/9810072;
           C.-S. Chu and P.-M. Ho, Nucl. Phys. {\bf B550} (1999)
           151, hep-th/9812219; C.-S. Chu and P.-M. Ho, Nucl. Phys.
           {\bf B568} (2000) 447, hep-th/9906192; V. Schomerus, 
           JHEP 9906 (1999) 030, hep-th/9903205.

\bibitem{witten} N. Seiberg and E. Witten, JHEP 9909 (1999) 032,
           hep-th/9908142.

\bibitem{uvir} S. Minwalla, M. Van Raamsdonk, N. Seiberg, hep-th/9912072;
           M. Hayakawa, Phys. Lett. {\bf B478} (2000) 394, hep-th/9912094;
           A. Matusis, L. Susskind, N. Toumbas, hep-th/0002075;
           M. Van Raamsdonk and N. Seiberg, JHEP 0003 (2000) 035,
           hep-th/0002186. 

\bibitem{chull} M.R. Douglas, C. Hull, JHEP 9802 (1998) 008, 
           hep-th/9711165.

\bibitem{fisch} W. Fischler, E. Gorbatov, A. Kashani-Poor,
         S. Paban, P. Pouliot and J. Gomis, JHEP 0005 (2000) 024, 
         hep-th/0002067; W. Fischler, E. Gorbatov, A. Kashani-Poor, R.
         McNees, S. Paban and P. Pouliot, JHEP 0006 (2000) 032, 
         hep-th/0003216.

\bibitem{verlinde} L. Randall and R. Sundrum, Phys. Rev. Lett. {\bf 83}
         (1999) 3370, hep-th/9905221; L. Randall and R. Sundrum, Phys. 
         Rev. Lett. {\bf 83} (1999) 4690, hep-th/9906064; 
         H. Verlinde, Nucl. Phys. {\bf B580} (2000) 264, hep-th/9906182.

\bibitem{mehen} A. Connes, M.R. Douglas, A. Schwarz, JHEP 9802 (1998)
         003, hep-th/9711162; F. Ardalan, H. Arfaei and M.M. Sheikh-Jabbari, 
         JHEP 9902 (1999) 016, hep-th/9810072; R. Blumenhagen, 
         L. G{\"o}rlich, 
         B. K{\"o}rs and D. L{\"u}st, Nucl. Phys. {\bf B582} (2000) 44,
         hep-th/0003024; J. Gomis, T. Mehen and M. Wise, JHEP 0008
         (2000) 029, hep-th/0006160; R. Blumenhagen, L. G{\"o}rlich, 
         B. K{\"o}rs and D. L{\"u}st, JHEP 0010 (2000) 006, hep-th/0007024;
         S. Nam, hep-th/0008083; W.-H. Huang, hep-th/0010160.

\bibitem{kek} N. Ishibashi, S. Iso, H. Kawai, and Y. Kitazawa
         Nucl. Phys. {\bf B583} (2000) 159, hep-th/0004038.

\bibitem{arcioni} G. Arcioni, J.L.F. Barbon, J. Gomis, M.A.
         Vazquez-Mozo, JHEP 0006 (2000) 038, hep-th/0004080.

\bibitem{suss}  M.M. Sheikh-Jabbari, Phys. Lett. {\bf B455} (1999) 
         129, hep-th/9901080; D. Bigatti and L. Susskind, Phys. 
         Rev. {\bf D62} (2000) 066004, hep-th/9908056; Z. Yin, Phys. 
         Lett. {\bf B466} (1999) 234, hep-th/9908152. 

\bibitem{liu} H. Liu and Michelson, Phys.Rev. {\bf D62} (2000) 
         066003, hep-th/0004013.

\bibitem{rajaraman} A. Rajaraman and M. Rozali, JEHP 0004 (2000)
         033, hep-th/0003227;  Y. Kiem and S. Lee, Nucl. Phys. {\bf B586} 
         (2000) 303, hep-th/0003145.
  
\bibitem{shenker} J. Gomis, M. Kleban, T. Mehen, M. Rangamani and S. Shenker, 
         JHEP 0008 (2000) 011, hep-th/0003215.

\bibitem{technique}  O. Andreev, Phys. Lett. {\bf B481} (2000) 125,
         hep-th/0001118;A. Bilal, C.-S. Chu and R. Russo, Nucl. Phys. 
         {\bf B582} (2000) 65, hep-th/0003180; C.-S. Chu, R. Russo, 
         S. Sciuto, Nucl. Phys. {\bf B585} (2000) 193, hep-th/0004183
         Y. Kiem, S. Lee and J. Park, hep-th/0008002, to appear in
         Nucl. Phys. {\bf B}.

\bibitem{recent} Y. Kiem, D.-H. Park and H.-T. Sato, hep-th/0011019.

\bibitem{atick} L. Dixon, D. Friedan, E. Martinec and S. Shenker, 
         Nucl. Phys. {\bf B282} (1987) 13; A.B. Zamolodchikov, 
         Nucl. Phys. {\bf B285} (1987) 481; J. Atick and E. Witten,
         Nucl. Phys. {\bf B310} (1988) 291-334.
 
\bibitem{haru} K. Roland and H.T. Sato, Nucl. Phys. {\bf B480} (1996) 99;
         Nucl. Phys. {\bf B515} (1998) 488.

\end{thebibliography}
\end{document}